\documentclass[prodmode,acmcsur]{acmsmall}

\def\firstfoot{}
\runningfoot{}

\usepackage[ruled]{algorithm2e}

\SetAlFnt{\small}
\SetAlCapFnt{\small}
\SetAlCapNameFnt{\small}
\SetAlCapHSkip{0pt}
\IncMargin{-\parindent}
\usepackage{tabularx,ragged2e}

\usepackage[show]{chato-notes}

\usepackage{caption}
\usepackage{booktabs}
\usepackage{paralist}
\PassOptionsToPackage{hyphens}{url}
\usepackage{breakurl}

\newcommand{\spara}[1]{\smallskip\noindent{\bf #1}}

\begin{document}

\markboth{Imran et al.}{Social Media in Mass Emergency}

\title{Processing Social Media Messages in Mass Emergency: A Survey}
\author{Muhammad Imran
\affil{Qatar Computing Research Institute}
Carlos Castillo
\affil{Qatar Computing Research Institute}
Fernando Diaz
\affil{Microsoft Research}
Sarah Vieweg
\affil{Qatar Computing Research Institute}
}

\begin{abstract}
Social media platforms provide active communication channels during
mass convergence and emergency events such as disasters caused by natural hazards.
As a result, first responders, decision makers, and the public can use
this information to gain insight into the situation as it unfolds. 
In particular, many social media messages communicated during emergencies convey timely, actionable information.
Processing social media messages to obtain such information, however, involves solving multiple challenges including: parsing brief and informal messages, handling information overload, and prioritizing different types of information found in messages. 
These challenges can be mapped to classical information processing operations such as filtering, classifying, ranking, aggregating, extracting, and summarizing.
We survey the state of the art regarding computational methods to process social media messages and highlight both their contributions and shortcomings. In addition, we examine their particularities, and methodically examine a series of key sub-problems ranging from the detection of events to the creation of actionable and useful summaries. %
Research thus far has to a large extent produced methods to extract situational awareness information from social media; in this survey, we cover these various approaches, and highlight their benefits and their shortcomings. We conclude with research challenges that go beyond situational awareness, and begin to look at supporting decision-making and coordinating emergency-response actions.
\end{abstract}


\terms{Design, Algorithms, Performance}



\maketitle

\section{Introduction}\label{sec:introduction}

Crisis situations such as disasters brought on by natural hazards present unique challenges to those who study them, creating conditions that call for particular research methods~\cite{stallings_2012_unique}. 
In this paper, we survey methods for studying disasters from the perspective of information processing and management, specifically methods for processing social media content. 


Crisis situations---particularly those with little to no warning (known as ``sudden onset crises")---generate a situation that is rife with questions, uncertainties, and the need to make quick decisions, often with minimal information. When it comes to information scarcity, research in recent years has uncovered the increasingly important role of social media communications in disaster situations, and shown that information broadcast via social media can enhance situational awareness during a crisis situation~\cite{vieweg_2012_thesis}.
However, social media communications during disasters are now so abundant that it is necessary to sift through hundreds of thousands, and even millions, of data points to find information that is most useful during a given event.

%
The goal of this survey is to provide computer science researchers and software developers with computational methods they can use to create tools for formal response agencies, humanitarian organizations, and other end users with a way to successfully identify, filter, and organize the overwhelming amount of social media data that are produced during any given crisis. Such tools can help stakeholders make time-critical---and potentially life-saving---decisions.

\subsection{Social Media During Crisis Situations}\label{subsec:motivation}

\spara{Brief History.}
The use of Internet technologies to gather and disperse information in disaster situations, as well as to communicate among stakeholders, dates back to the late 1990s. Internet historians point to online newsgroups and email clients that were used to coordinate protests in Indonesia in 1998~\cite{poole2005internet}. In addition, there are cases of websites being set up in response to crises in 2003~\cite{palen2007citizen}. 
To the best of our knowledge, 2004 is the first year in which a user-generated content website was used in response to a crisis; after the Indian Ocean Tsunami of December 26 that year, an electronic bulletin board was set up and moderated for 10 days.\footnote{\url{http://www.thefreelibrary.com/www.p-h-u-k-e-t.com+Has+Served+Its+Purpose+After+the+Tsunami\%3B+Site...-a0126803919}}  
In addition, in the aftermath of Hurricane Katrina, which struck the city of New Orleans in the United States in 2005, significant emergency response activity took place on MySpace~\cite{shklovski_2010_katrina}. 
%
%
One of the earliest known cases of people using microblogging service Twitter in an emergency was during severe wildfires that took place near San Diego, California (in the United States) in 2007.\footnote{Eric Frost, personal communication.} Since then, it has become common practice for affected populations and concerned others to use Twitter to communicate, ask questions, collect and spread information, and organize response efforts (among other tasks)~\cite{starbird2010chatter,Vieweg:2010,sarcevic2012beacons,starbird2013delivering,imran2014aidr,cobb2014designing}. 

\spara{Today.}
The growing adoption of social media during disasters has created opportunities for information propagation that would not exist otherwise. Emergency response agencies routinely post information such as emergency alerts and advice through these channels,\footnote{See e.g. \url{https://blog.twitter.com/2013/twitter-alerts-critical-information-when-you-need-it-most}} but social media enables much more than ``top-down'' communications.
People post situation-sensitive information on social media related to what they experience, witness, and/or hear from other sources~\cite{hughes2009twitter}. This practice allows both affected populations and those outside the impact zone to learn about the situation \textit{first hand} and in near real-time. 

We know that information posted to social media platforms in time- and safety-critical circumstances can be of great value to those tasked with making decisions in these fraught situations. Previous research has shown that information which contributes to situational awareness is reported via Twitter (and other social media platforms) during mass emergencies~\cite{Vieweg:2010,vieweg_2012_thesis,imran2014aidr}. Now, those tasked with formal response efforts---from local fire departments\footnote{\url{http://edition.cnn.com/2012/11/01/tech/social-media/twitter-fdny/}} to international aid agencies---are working to incorporate information broadcast on social media platforms into their processes and procedures. Many emergency responders and humanitarian officials recognize the value of the information posted on social media platforms by members of the public (and others), and are interested in finding ways to quickly and easily locate and organize that information that is of most use to them~\cite{hughes_2012_phd}.\footnote{Andrej Verity, personal communication}
Some agencies have even begun to formally incorporate social media monitoring and communication during mass emergency situations. The American Red Cross (ARC), in a survey, reported the effectiveness of social media and mobile apps.\footnote{\url{http://www.redcross.org/news/press-release/More-Americans-Using-Mobile-Apps-in-Emergencies}} ARC recently opened their Social Media Digital Operations Center for Humanitarian Relief. The goals of the center are to ``source additional information from affected areas during emergencies to better serve those who need help; spot trends and better anticipate the public's needs; and connect people with the resources they need, like food, water, shelter or even emotional support."\footnote{http://www.redcross.org/news/press-release/The-American-Red-Cross-and-Dell-Launch-First-Of-Its-Kind-Social-Media-Digital-Operations-Center-for-Humanitarian-Relief} 
Though the ARC is currently one of the few (possibly the only) formal agencies to support such a center, it is likely that similar operations will begin within other organizations. Among other similar examples, the Australia Crisis Tracker,\footnote{\url{http://www.crisistracker.com.au/}} which is a machine-learning based tool to filter spam and categorize data into different event types, is also deployed with the Australian Red Cross.\footnote{\url{http://www.research.ibm.com/articles/crisis-tracker.shtml}} 

Though formal response agencies express interest in incorporating social media into their processes, obstacles exist. For example, a recent survey by the US Congressional Research Service cites administrative cost as a significant barrier to adopting social media during emergencies:
``The number of personnel required to monitor multiple social media sources, verify the accuracy of incoming information, and respond to and redirect incoming messages is also uncertain ... Responding to each message in a timely manner could be time consuming and might require an increase in the number of employees ...''~\cite{lindsay_2011_social_media_disasters}
Others have expressed concerns including issues related to roles, responsibilities, and liabilities; difficulties evaluating the veracity, trustworthiness, and reliability of information; and information overload in general~\cite{vieweg_2014_integrating,hughes_2014_smem}. 
Computational methods can help overcome some of these obstacles, by reducing the amount of information to be examined by humans. Automatic methods are necessary when human computation is limited, and in the following sections, we detail what those methods entail.

\subsection{Background Reading}\label{subsec:background-readings}

Sociologists  began researching human behavior in mass emergency situations long before the Internet, or even modern computing.
The purpose of this section is not to provide the reader with an exhaustive list of sociology of disaster literature; we highlight a few foundational readings that are helpful for the computer science, information science, technology, and social media scholars to gain quick insight into the rich and varied field of sociology of disaster. 

E.L. Quarantelli's 2002 chapter ``The Disaster Research Center (DRC) Field Studies of Organized Behavior in the Crisis Time Period of Disasters'' (in \textit{Methods of Disaster Research} edited by R.A. Stallings~\shortcite{quarantelli_2002_drc}) provides a brief history of one of the foremost disaster research institutes in the United States. Quarantelli gives background on the Disaster Research Center, and explains the strategic as well as academically-oriented decisions that were made in order to highlight the importance of studying the social science aspects of disaster. 

In his edited volume ``Disasters by Design,'' Dennis S. Mileti~\shortcite{mileti_1999_disasters} and the contributing authors aim to reach a general (i.e. non-academic) audience and provide background on disasters caused by natural hazards. The volume is comprised of ``synthesized statements of what is known, collectively, about hazards and human coping strategies.'' Mileti and colleagues point to causes of disaster, which happen when three major systems---the physical, social, and built environments---interact in complex ways. The authors' goal is to give the reader a way to understand how to study disaster situations, with a final goal of helping members of the public create more resilient communities. 

When it comes to combining studies of disaster with the use of information technology, including social media, the Harvard Humanitarian Initiative~\shortcite{hhe_2011_disaster2} presents an in-depth analysis of the response to the earthquake in Haiti in January 2010.
With a stronger focus in social media, a recent survey by Hughes, Peterson and Palen considers the motivating factors of emergency responders regarding their use of social media data. The authors describe the challenges they face, best practices regarding the adoption of social media by formal response organizations, and also touch on instances of integrated, end-to-end systems that are currently being built to meet these needs~\cite{hughes_2014_smem}. In addition, an article by Palen and Liu~\shortcite{palen2007citizen} was one of the first to provide an early assessment regarding how information and communication technology can support the participation of the public during crisis situations. Since then, many articles that focus on the role of social media in disaster have been published, but the two we mention here provide a good ``first glance'' to readers who are new to the field.

Our brief overview of foundational reading would not be complete without mentioning the much-discussed issue of trust and the use of social media. A recent ACM Computing Survey looks at this very topic ~\cite{Sherchan:2013:STS:2501654.2501661}. The authors review the various definitions of ``trust'' from a variety of academic disciplines, discuss the factors that contribute to notions of trust, and combine the complex and much-scrutinized idea of trust with computing and social network research. 

\subsection{Scope and Organization}

The overarching problem we aim to confront in this article is that of extracting time-critical information from social media that is useful for emergency responders, affected communities, and other concerned populations in disaster situations.

We note that social media analysis has been used for a number of applications in the domains of economics, politics (e.g.~\citeN{hong2011does}) and public health (e.g.~\citeN{aramaki2011twitter}); for a survey, see~\citeN{cikm_2013_twitterreal}. Even if we consider only applications to time-critical settings, we note that often the same methods described in this survey can be applied to the analysis of social media during mass converge events, such as large political conventions, concerts, or sports events; or for monitoring the performance of a media campaign or a televised debate, among similar applications. However, while many of the methods and algorithms that we describe can be used for other purposes, we explain them from the perspective of their applications to mass emergencies, a topic that has a specific scientific and technical community, and that targets a particular set of use-cases.

The following two sections briefly describe our target end-user audience, and their information needs (Section~\ref{sec:users}), and end-to-end integrated systems (Section~\ref{sec:systems}). The subsequent sections form the main technical part of this survey and present a systematic analysis of the computational methods we cover.

\begin{itemize}
 \item Section~\ref{sec:data} starts with a general characterization of social media messages broadcast during disasters. Next, it introduces methods for programmatically acquiring data from social media, and for pre-processing the data.
 \item Section~\ref{sec:event-detection} covers methods for the detection of new events, which involves detecting the first message on a given topic or sub-topic. The section also covers how to track these events, i.e. how to collect further messages belonging to the same topic.
 \item Section~\ref{sec:finding_information} outlines methods to mine and aggregate information. These methods include unsupervised classification (and clustering), supervised classification, information extraction, and summarization.
 \item Section~\ref{sec:semantics} presents how semantic technologies can be applied in this domain. This corresponds, first, to enrich the content with semantic information, and second, to use an ontology for disaster management to describe the content of the messages.
\end{itemize}

The final section concludes the survey, and outlines current research directions.

\section{Users and Information Needs}\label{sec:users}

Much of the research we present here focuses on the computational aspects of processing social media messages in time- and safety-critical situations. It is additionally important to consider the end users of these technological solutions; those who benefit from having curated information that describes a disaster or crisis and enhances situational awareness, including formal response agencies, humanitarian organizations, and members of the public.

\subsection{Public Participation in Crises}

Ideally, to understand how the public participates in social media during crises, we should start by asking how the public reacts in general to crises.

Contrary to Hollywood renditions of disaster situations, human response to crises is not one of panic and mayhem~\cite{mitchell2000catastrophe}. Victims of disaster do not lose control, run amok, nor flee the area in fear. Instead, they make quick decisions based on the information available to them at the time, which often allow them to save their own lives as well as help those around them~\cite{mitchell2000catastrophe}.
Neighbors, friends, and other members of the public are the first to respond when a disaster strikes. They rush to the scene to perform search and rescue operations, administer first aid, and perform critical tasks necessary in the first moments of response. Often, these ``first responders'' are victims of the disaster themselves~\cite{dynes1970organized}.
The role of the public in disaster response efforts is critical, and with the growing use of social media to gather and disperse information, organize relief efforts, and communicate, those members of the public who can play a valuable role in these situations is no longer limited to those in the area of impact.

As~\citeN{dynes1994community} explains, emergencies do not render victims incapable of helping themselves and others, nor create a situation in which they are unable to make intelligent, personally meaningful decisions. What emergencies \textit{do} create is an environment in which new and perhaps unexpected problems are presented, which members of the public are called upon to solve.
Research in recent years on the use of social media in disasters shows how members of the public, formal response agencies, and other stakeholders have taken to online outlets to perform tasks such as
communicating about hospital availability~\cite{starbird2013delivering},
coordinating medical responses~\cite{sarcevic2012beacons},
and communicating with the public during various crises~\cite{cobb2014designing}, among many others.
%
%
These users interact in complex ways including producing, distributing and organizing content~\cite{starbird2010chatter}.


\subsection{Differences in Information Needs}

The recognition that social media communications are a valid and useful source of information throughout the disaster lifecycle (preparation, impact, response, and recovery) is increasing among the many stakeholders who take action in disaster situations. In particular, members of the public, formal response agencies, and local, national and international aid organizations are all aware of the ability to use social media to gather and disperse timely information in the aftermath of disaster, but the specific information they seek---and their ability to put it to use---may differ~\cite{vieweg_2012_thesis}.

Depending on the circumstances of the disaster, and what roles and duties the various stakeholders are responsible for, their specific information needs will vary. For example, in a wildfire situation that affects a community, members of a formal response organization such as local police or area firefighters can benefit from information such as where people are smelling smoke, what precautions they are taking (e.g. clearing brush, watering yards), and what traffic patterns look like. In a large-scale, sudden-onset disaster such as a typhoon or earthquake, humanitarian agencies, such as the various branches of the United Nations, 
benefit from information that details the current situation ``on the ground,'' such as where electricity has been disabled, or where people are without food and water.
In any disaster situation, members of the public play a variety of roles and take on many tasks; the information they find valuable may be very personal---i.e. hearing that a friend or loved one is safe, or it may be more broadly applicable, such as the status of a certain neighborhood or town. 

Overall, the information any individual, group, or organization finds useful and seeks out in a disaster will depend upon their goals. Is it a group interested in providing food to children? Is it an organization that can set up a field hospital? Is it an individual living in a foreign country who is concerned about her or his family? The different types of information sought by these different stakeholders may be broadcast on Twitter, but to find it quickly, users rely on technological methods to sift through the millions of tweets broadcast at any given time to find useful information. Further information and a deeper perspective on users of social media in disaster can be found in \citeN{hughes_2014_smem} and \citeN{hughes2012evolving}.

\section{Systems for Crisis-Related Social Media Monitoring} \label{sec:systems}

Table~\ref{tbl:systems} provides examples of existing systems described in the literature that extract crisis-relevant information from social media.\footnote{The list is not extensive, and does not include tools such as Radian6 (\url{http://www.salesforcemarketingcloud.com/products/social-media-listening/}) that have not been described in the literature, but might be relevant for other reasons---e.g. in the case of Radian6, because it is used by the American Red Cross.}
The systems we include have varying degrees of maturity; some have been deployed in real-life situations, while others remain under development.

\begin{table}[t]
\caption{Example systems described in the academic literature that extract crisis-relevant information from social media.}
\label{tbl:systems}
\scriptsize\begin{tabular}{p{3.1in}l}\toprule
System name \\
~~ Data; example capabilities & Reference and URL \\\midrule
\textit{Twitris} & \cite{sheth2010understanding,purohit2013twitris} \\
~~ Twitter; semantic enrichment, classify automatically, geotag & \url{http://twitris.knoesis.org/} \\
\textit{SensePlace2} & \cite{maceachren2011senseplace2} \\
~~ Twitter; geotag, visualize heat-maps based on geotags & \url{http://www.geovista.psu.edu/SensePlace2/} \\[2pt]
\textit{EMERSE}: Enhanced Messaging for the Emergency Response Sector & \cite{caragea2011classifying} \\
~~ Twitter and SMS; machine-translate, classify automatically, alerts  & \url{http://emerse.ist.psu.edu/}\\[2pt]
\textit{ESA}: Emergency Situation Awareness & \cite{yin2012using,power2014emergency}\\
~~ Twitter; detect bursts, classify, cluster, geotag & \url{https://esa.csiro.au/} \\[2pt]
\textit{Twitcident} & \cite{abel2012semantics}\\
~~ Twitter and TwitPic; semantic enrichment, classify & \url{http://wis.ewi.tudelft.nl/twitcident/} \\[2pt]
\textit{CrisisTracker} & \cite{rogstadius2013crisistracker}\\
~~ Twitter; cluster, annotate manually & \url{https://github.com/jakobrogstadius/crisistracker}\\[2pt]
\textit{Tweedr} & \cite{ashktorab2014tweedr}\\ 
~~ Twitter; classify automatically, extract information, geotag & \url{https://github.com/dssg/tweedr} \\[2pt] 
\textit{AIDR}: Artificial Intelligence for Disaster Response & \cite{imran2014aidr}\\
~~ Twitter; annotate manually, classify automatically & \url{http://aidr.qcri.org/} \\[2pt]
\bottomrule
\end{tabular}
\end{table}

Most existing systems are built around the concept of a \textit{dashboard}, or a set of visual displays that provides a summary of social media during the crisis according to temporal, spatial, and thematic aspects. Common elements in these displays include: 

\begin{itemize}
 \item Lists/timelines showing recent or important messages, sometimes grouping the messages into clusters or categories.
 
 \item Time series graphs representing the volume of a hashtag, word, phrase, or concept over time, and sometimes marking peaks of activity.
 
 \item Maps including geotagged messages or interpolated regions, possibly layered according to different topics.
 
 \item Pie charts or other visual summaries of the proportion of different messages.
\end{itemize}

These visual elements are powered by computational capabilities that include:

\begin{itemize}

 \item Collections of social media messages matching a given criterion, from one or multiple social media and/or Short Message Service (SMS) streams, typically with a focus on Twitter (described in Section~\ref{sec:data}).
 
 \item Natural Language Processing (NLP), including Named Entity Recognition (NER) and linking of named entities to concepts (described in Section~\ref{sec:data-preprocess}).
 
 \item Extraction of information from the messages, including geotagging (described in Section~\ref{sec:geotag}).
 
 \item Monitoring the volume of messages (or sets of messages) to detect or help detect sub-events within a crisis, and sometimes the crisis itself (described in Section~\ref{sec:event-detection}), possibly including the generation of user-defined alerts when certain conditions are met.
 
 \item Clustering or automatic grouping of similar messages (described in Section~\ref{sec:unsupervised_classification}).
 
 \item Classification of messages or groups of messages manually or automatically (see Section~\ref{sec:supervised_classification}).
 
 \item Automatic translation of messages.
\end{itemize}

Though some of these systems are based on input or feedback provided by emergency responders and other officials, we note that to a large extent they are framed as a way to process social media data during crisis situations; their goal is not to address specific needs of emergency responders or other stakeholders.
This focus on \textit{processing} social media data possibly impacts the adoption of these systems by the practitioner community. Methodologies such as participatory design have been proposed to improve the matching between e.g. the needs of public information officers during a crisis, and the tools built by researchers and developers~\cite{hughes_2014_pio}.

When the goal of meeting the specific need of users \textit{is} stated explicitly in the design of systems, it often revolves around \textit{enhancing situational awareness}, defined in~\cite{endsley1995toward} as ``the perception of elements in the environment within a volume of time and space, the comprehension of their meaning, and the projection of their status in the near future.''

For instance, \textit{ESA}~\cite{yin2012using,power2014emergency} aims at enhancing situational awareness with respect to crises induced by natural hazards, particularly earthquakes. This is done by presenting information in time (frequency series) and space (maps), which is achieved by performing event detection, text classification, online clustering and geotagging. 
Similarly, \textit{SensePlace2}~\cite{maceachren2011senseplace2} is presented as a \textit{geovisual analytics} system, which filters and extracts geographical, temporal and thematic information from tweets in order to present them in a layered map.
Data from social media can also be presented along with data from physical sensors, for instance to overlay earthquakes detected by seismic sensors on a map presenting social media data~\cite{avvenuti2014ears,musaevlitmus}.

In parallel to approaches that use Natural Language Processing (NLP) techniques to enhance situation awareness, \emph{Crisis Mapping} emerged as an alternative type of system, by employing digital volunteers to collect, classify, and geotag messages~\cite{okolloh2009ushahidi,meier_2015_dh}, and eventually by using the input from those volunteers to train machines to perform these tasks automatically~\cite{imran2014aidr}.

\section{Data Characterization, Acquisition, and Preparation}\label{sec:data}
Both academics and practitioners gather social media data during crisis events.  In this section, we describe the common practices used to collect, represent, and process these data.

\subsection{Characteristics of Messages Broadcast on Social Media in Disaster}

Social media is a general term that encompasses a variety of platforms on which user-generated content can be disseminated and consumed, and where users can connect with others. This definition currently includes blogging and micro-blogging, social networking sites, social media sharing platforms, and {\em wikis}~\cite{vandijck2013social}.

Activities such as staying in touch with friends and family, and connecting with others, have driven the growth of social media platforms.\footnote{\url{http://www.pewinternet.org/2011/11/15/why-americans-use-social-media/}}
Currently, different social networking sites are used for different purposes, but commonalities do exist. For instance, the top 3 activities on Twitter are to (1) post about daily activities, (2) upload and share photos, and (3) comment on posts of others; while on Facebook they are to (1) upload and share photos, (2) message with friends on a one-on-one basis, and (3) comment on posts of friends~\cite{globalwebindex2013q2}.

In any of these platforms, an increase in social media communications can be triggered by a variety of causes, which can be divided into \textit{endogenous} and \textit{exogenous}~\cite{crane2008robust}. Endogenous causes refer to phenomena in which an idea or ``meme'' gains popularity by a process of {\em viral contagion} or {\em information cascade}, where content spreads rapidly through a network, potentially reaching a significant fraction of all the users~\cite{chen_2014_propagation}.

Exogenous causes refer to large-scale events, usually happening in the physical world, of wide interest to social media users. 
Emergencies and mass convergence events are examples of an exogenous cause, and during such events, we know that communications activity increases. 
For instance, it has been observed that mobile network usage---both in terms of phone calls and SMS---increases in emergency situations~\cite{gao_2014_quantifying}.
The same is true for social media usage, which ``rises during disasters as people seek immediate and in-depth information''~\cite{fraustino_2012_start}.
\footnote{For instance, activity on Facebook was observed to increase significantly in the areas most affected by the August 2014 earthquake in California, USA \url{https://www.facebook.com/notes/facebook-data-science/on-facebook-when-the-earth-shakes/10152488877538859}}

To illustrate the types of information that affected populations broadcast specifically on the popular microblogging platform Twitter, we turn to some example messages that have been highlighted in previous literature:

\begin{itemize}
 \item ``OMG! The fire seems out of control: It's running down the hills!'' (bush fire near Marseilles, France, in 2009,  quoted from Twitter in \citeN{longueville2009omg})
 \item ``Red River at East Grand Forks is 48.70 feet, +20.7 feet of flood stage, -5.65 feet of 1997 crest. \#flood09'' (automatically-generated tweet during Red River Valley floods in 2009, quoted from Twitter in~\citeN{starbird2010chatter})
 \item ``Anyone know of volunteer opportunities for hurricane Sandy? Would like to try and help in any way possible'' (Hurricane Sandy 2013, quoted from Twitter in~\citeN{purohit2013emergency})
 \item ``My moms backyard in Hatteras. That dock is usually about 3 feet above water [photo]'' (Hurricane Sandy 2013, quoted from Reddit in~\citeN{leavitt2014upvoting})
 \item ``Sirens going off now!! Take cover...be safe!'' (Moore Tornado 2013, quoted from Twitter in~\citeN{blanford2014tweeting})
\end{itemize}

Though the above are only a few examples, they convey a sense of the types of information posted during an event, and show that it is varied.
\citeN{vieweg_2012_thesis} points to this variation in her research that is based on a detailed study of four crisis events, in which she identifies in Twitter 35 types of information in three broad categories defined by~\citeN{mileti_1999_disasters}: social environment, built environment, and physical environment. She points out that social environment messages describe anything having to do with people and their reactions to the crisis, built environment messages correspond to information and updates about property and infrastructure, and physical environment messages include updates about the hazard agent, weather, and other environmental factors (see Section~\ref{subsec:content-categories} for details on different ways of categorizing this information).

Quantifying the amount of information found in social media based on type is even more difficult than locating that information in the first place. Important variations have been observed across crises (even for similar events) and at different stages of a crisis~\cite{blanford2014tweeting,imran2013practical}.
\citeN{olteanu2014crisislex} looked at the prevalence of three broad categories of information in tweets related to six crisis events. The results show large variabilities in the number of tweets reporting the negative consequences of an event (20\%-60\%), those offering or asking for donations (15\%-70\%) and those warning about risks or providing advice (5\%-20\%).


%
Using tweets as an indication of spatial zones during a disaster is also possible. For instance, Acar and Muraki~\citeyear{acar2011twitter} examine the use of Twitter during an earthquake in Japan and observed that tweets from affected areas include more requests for help and more warnings, while tweets from other areas which are far from the disaster epicenter tend to mostly include other types of information, such as concern and condolences.



\subsection{Data Acquisition}\label{sec:data-acquisition}


Most large social media platforms provide programmatic access to their content through an Application Programming Interface (API). However, the details of these APIs vary substantially from one platform to another, and also change over time.

APIs to access social media data typically belong to one of two types: those allowing to query an archive of past messages (also known as {\em search} APIs), and those allowing data collectors to subscribe to a real-time data feed (also known as {\em streaming} or {\em filtering} APIs). 
Both types of APIs typically allow data collectors to express an information need, including one or several of the following constraints: (i) a time period, (ii) a geographical region for messages that have GPS coordinates (which are currently the minority), (iii) a set of keywords that must be present in the messages, which requires the use of a query language whose expressiveness varies across platforms.
In the case of archive/search APIs, messages are returned sorted by relevance (a combination of several factors, including recency), or just by recency. In the case of real-time/streaming/filtering APIs, messages are returned in order of their posting time.

Data collection strategies impact the data obtained and analytic results. For instance, selecting messages in the geographical region affected by a disaster vs. selecting messages based on a keyword-based query may return datasets having different characteristics~\cite{olteanu2014crisislex}.

\spara{Data availability.} \urldef{\weiboapi}\url{http://open.weibo.com/wiki/API%E6%96%87%E6%A1%A3/en}
In addition to issues regarding network connectivity during disasters~\cite{jennex2012social}, access to social media data is in general limited, which is a serious obstacle to research and development in this space~\cite{reuter2014technical}.

First, historical, archived social media can typically only be queried through {\em search APIs}, which are limited in terms of number of queries or data requested per unit of time. The exceptions are (i) a 1\% data sample collected from Twitter by the Internet Archive,\footnote{Available at \url{https://archive.org/details/twitterstream}} and (ii) collections created by researchers using automatic data collection methods (e.g. web crawling), which are against the terms of services of most social media platforms.

Second, access to real-time data is quite limited. Twitter offers a public streaming API providing a random sample of 1\% of all postings, plus the possibility of filtering all public postings by keywords.\footnote{Available at \url{https://dev.twitter.com/docs/api/streaming}}
This is in contrast with most large social media platforms, which do not offer this level of data access publicly.\footnote{As of November 2014, we found no publicly-available equivalent of Twitter's {\em streaming API} in Sina Weibo, Facebook, YouTube, Google Plus, or Tumblr.}

The consequence of these limitations is that most work on crisis-related messages is done using Twitter data (with few exceptions e.g.~Facebook~\cite{bird2012flooding} and Reddit~\cite{leavitt2014upvoting}), which provides an incomplete view of a crisis situation, as there are many different online social media sites that might be used for different purposes. Limitations in the amount of data that can be collected and in general a dependency on a small set of data providers or APIs, further reduces the efficiency and effectiveness of tools for handling crisis-related social media messages.

\subsection{Data Pre-Processing}\label{sec:data-preprocess}

Most researchers and practitioners prepare social media data by {\em pre-processing} it, using some of the methods outlined below, before performing the actual analysis.
Many pre-processing techniques are available, and the choice depends on the type of data at hand and the goals of the analysis.

\spara{Natural Language Processing (NLP).}
The text of the messages can be pre-processed by using an NLP toolkit. Typical operations include tokenization, part-of-speech tagging (POS), semantic role labeling, dependency parsing, named entity recognition and entity linking. A number of off-the-shelf implementations of these operations are available online, e.g. the Stanford NLP\footnote{\url{http://www-nlp.stanford.edu/software/}} or NLTK for Python.\footnote{\url{http://www.nltk.org/}}
Social media-specific NLP toolkits can also be used. For instance, ArkNLP~\cite{owoputi2013improved}, which is trained on Twitter data, is able to recognize Internet idioms such as ``ikr'' (\emph{I know, right?}) and assign them the correct Part Of Speech (POS) tag (interjection, in this case).
Additionally, higher-level operations can be applied, including applying sentiment analysis methods to infer aspects of the emotion conveyed by a piece of text~\cite{pang2008opinion}.

\spara{Feature extraction.}\label{subsec:feature-extraction}
For many automatic information processing algorithms (e.g. machine learning), each data item must be represented accurately as an information record. The representation of choice for text is typically a numerical vector in which each position corresponds to a word or phrase---this is known as the vector space model in information retrieval. The value in each position can be a binary variable, indicating the presence or absence of that word or phrase in the message, or a number following some weighting scheme, such as TF-IDF, which favors terms that are infrequent in the collection~\cite{mir2ed}. 
To avoid having too many variables, textual features can be discarded by removing stopwords and functional words, or by normalizing words using stemming or lemmatization (e.g. considering ``damaging'' and ``damage'' as equivalent), or other means.

Additionally, other text-based features can be added, such as the length of the text in words or characters, and the number of question or exclamation marks. If some NLP pre-processing is performed on the text---such as part-of-speech-tagging---a feature such as \emph{noun:fire} (instead of \emph{verb:fire}) can be used to distinguish that the word ``fire'' is being used in a message as a noun (``I heard a fire alarm'') instead of a verb (``They should fire him''). 

In the case of tweets, characteristics such as the presence of user mentions (``@user''), URLs, or hashtags can be included as features.
In the case of images or video, content-based features such as colors, textures and shapes can be included~(see e.g. \citeN{liu2007survey} for a survey). Additionally, features such as the date of a message, tags associated with it, the number of views/comments it has received, or information about its author, are often available in a platform-dependent manner.

Obviously, one can spend a great deal of time constructing features by hand.  In order to guide this exploration, both researchers and practitioners should prioritize the development and understanding of features likely to be correlated with the target variable (e.g. tweet classification).  By the same token, features valuable to one target variable may not be important at all for a different target variable.  Although often under-appreciated, feature engineering is perhaps the most important part of a modeling exercise.  

\spara{De-duplication.}
Further reduction of the amount of data to be processed can be achieved by removing near-duplicate messages. Given that in many social media platforms the number of people re-posting a message can be interpreted as a measure of its importance, whenever removing near-duplicates it is advisable to save the number of near-duplicates that have been found (e.g. as done in~\citeN{rogstadius2013crisistracker} to prioritize highly-reposted stories). De-duplication can be done by applying a clustering method (see Section~\ref{sec:unsupervised_classification}).

\spara{Filtering.}
A fraction of the messages collected will not be relevant for a given crisis. This fraction depends on the specific collection method used, as discussed by~\citeN{olteanu2014crisislex}, and on other factors, such as the presence of off-topic messages using the same tags or keywords as the on-topic ones~\cite{qu2011microblogging}. These messages can be post-filtered using human labeling or crowdsourcing, keyword-based heuristics, or automatic classification.

Additionally, many messages are posted automatically on social media for financial gain, exploiting the attention that a certain hashtag has received. These unsolicited commercial messages are known as {\em spam}~\cite{gupta2012credibility,uddinunderstanding} and there are well-studied methods that can remove a substantial portion of them~\cite{benevenuto2010detecting}.
Finally, in some cases we might want to also remove messages posted by automatic agents or social media {\em bots}. Their identification is similar to that of spammers.

\subsection{Geotagging and geocoding}\label{sec:geotag} 

Attaching geographical coordinates to a message (a process known as {\em geotagging}) is useful for a number of tasks in disaster response~\cite{graham:geolocation,ikawa_2013_location,lingad:www2013}.  
Geotagging allows the retrieval of information about a {\em local event}, by filtering the messages corresponding to a particular geographical region.
It allows the visualization of information about an event on a map, possibly making it more actionable for emergency responders. 
Geotagging can also be used for higher-level tasks, such as helping predict epidemic transmission of diseases based on geographical proximity~\cite{brennan2013towards}. 

The availability of machine-readable location information in social media messages, in the form of metadata, depends on the user's device having the capacity to know its location (e.g. via Global Positioning System (GPS)), on the specific client software having the capability to read this from the device, and most importantly, on the user enabling this feature explicitly (opting-in). In practice, only 2\% of crisis-related messages include machine-readable location information \cite{burton:twitter-geolocation}.

However, while explicit metadata about locations may be absent, many messages in social media do contain implicit references to names of places (e.g. ``The Christchurch hospital is operational''). 
{\em Geocoding} refers to finding these geographical references in the text, and linking them to geographical coordinates.
This can be done by using a named entity extractor to extract potential candidates, and then comparing those candidates with a list of place names. This is the approach used by e.g.~\citeN{maceachren2011senseplace2} which uses Gate\footnote{\url{https://gate.ac.uk/}} for the first task and Geonames\footnote{\url{http://geonames.org/}} for the second.

While building a comprehensive database of geospatial information, including place names, is an important component of geotagging (see e.g.~\citeN{middleton2014real}), geotagging is not merely a dictionary look-up process because of ambiguities. These ambiguities are known as ``{\em geo/non-geo}'' and ``{\em geo/geo}.''
A geo/non-geo ambiguity occurs for instance in the message ``Let's play Texas Hold 'em,'' that does not refer to the state of Texas in the USA. A geo/geo ambiguity is found in the message ``There is a fire in Paris,'' which may refer to the capital of France, or to any of more than a dozen places on Earth sharing the same name. For instance, in~\citeN{sultanik2012rapid}, authors proposed an unsupervised approach to extract and disambiguate location mentions in Twitter messages during crisis situations.

In general, geotagging is done through probabilistic methods (see e.g.~\citeN{cheng_2010_where}), often exploiting contextual clues. These clues may include the general location of a crisis, information about nearby places, and location information indicated by users in their profiles~\cite{gelernter2011geo}.

\subsection{Archived versus Live Data Processing}\label{sec:retro-vs-online}

Depending on the urgency with which the output of an analysis is required, data may be provided to an algorithm either as an archive, for \textit{retrospective} analysis, or as a live data feed, for \textit{real-time} analysis. These correspond to two standard concepts in computer science: off-line processing and on-line processing.

Retrospective data analysis (\textit{off-line processing}) starts with a batch of data relevant to an event, usually containing messages over the entire time range of interest. For example, we might consider re-constructing a timeline of events in the aftermath of an earthquake, by looking at all of the tweets from the moment of the earthquake up to two weeks after. In deciding how to build the timeline, we have the complete context of events during this two week window. 

Live data analysis (\textit{on-line processing}) is done over a stream of data relevant to the event, usually provided in real-time or with a short delay. For example, we might consider constructing a timeline of the events in the aftermath of an earthquake \textit{as we observe new tweets}; in deciding how to build the timeline, we have an incomplete context of the events and their future repercussions.

It is possible for algorithms to lie in between, operating on small batches of data at regular intervals (e.g. hourly, daily). 
The trade-off between retrospective and live data is a matter of accuracy versus latency.
Retrospective data analysis maximizes our context and, as a result, gives us an accurate picture of the data. However, because we have to wait for the data to accumulate, we incur latency between when an event happens and when it is processed.
Live data analysis, on the other hand, minimizes the latency but, because we have partial information, we may incur lower accuracy.
The choice of collection methodology depends on the use case. Crisis responders may want lower latency in order to better respond to a developing situation; forensic analysts may want higher accuracy and have the benefit of waiting for data to be collected.

While developing an algorithm, we can use retrospective data to \textit{simulate} live data. This is a standard experimental methodology that has been used in the past for information filtering tasks~\cite{voorhees:trecbook} and, more recently, for crisis informatics~\cite{qi:temporal-summarization,trec2013:temporal-summarization-overview}.

\subsection{Challenges} \label{sec:data-challenges}

There are a number of challenges associated with the processing of social media messages. In this section, we group them into two high-level categories: scalability and content. We defer specific challenges (e.g. challenges to event detection) to their respective technical sections.

\spara{Scalability issues.}
Large crises often generate an explosion of social media activity.  
Data size may be an issue, as for crises that last several days, millions of messages may be recorded. While the text of each message can be sorted, a data record for e.g. a Twitter message (140 characters of text) is around 4KB when we consider the metadata attached to each message. Thus, a Twitter collection for a crisis is then on the order of several hundred megabytes to a few gigabytes. In addition, multimedia objects such as images and videos may significantly increase the storage space requirements.

Data velocity may be a more challenging issue, especially considering that data does not flow at a constant rate but experiences drastic variations. The largest documented peak of tweets per minute during a natural hazard that we are aware of is 16,000 tweets per minute.\footnote{During Hurricane Sandy in 2012: http://www.cbsnews.com/news/social-media-a-news-source-and-tool-during-superstorm-sandy/} 

Finally, redundancy, which is commonly cited as a scalability challenge, to some extent cannot be avoided in this setting. Repeated (re-posted/re-tweeted) messages are common in time-sensitive social media, even encouraged, as in some platforms messages that gain more notoriety are those that are simply repeated more.



\spara{Content issues.}\label{subsec:content-issues}
Microblog messages are brief and informal. In addition, this type of messaging is often seen by users to be more akin to speech, as opposed to a form of writing, which---compounded with technological, cross-lingual and cross-cultural factors---implies that on the Internet ``we find language that is fragmentary, laden with typographical errors, often bereft of punctuation, and sometimes downright incoherent''~\cite{baron2003language}. 
This poses significant challenges to computational methods, and can lead to poor and misleading results by what is known as the ``garbage in, garbage out'' principle. 

Messages are also highly heterogeneous, with multiple sources (e.g. traditional media sources, eyewitness accounts, etc.), and varying levels of quality. Quality itself is an important and complex question for crisis managers, and encompasses a number of attributes including objectivity, clarity, timeliness and conciseness, among others~\cite{friberg2011information}.
Additionally, different languages can be present in the same crisis and sometimes in the same message---a phenomena known as ``borrowing,'' and ``code switching.'' This makes it difficult for both machines and humans (e.g. content annotators) to understand or classify messages.

Finally, brief messages sent during a crisis often assume a shared context from which only a minor part is sometimes made explicit. The area of study in linguistics known as {\em pragmatics} focuses on ``communication in context,'' and explains how people are able to infer the meaning of the communications because humans are very adept at understanding context. So,  in the case of Twitter communications, a reader can understand the tweet author's intent because she or he knows the context within which that tweet is being broadcast. Current computational methods are not able to make the same inferences humans do, and thus cannot achieve the same level of understanding~\cite{vieweg2014pragmatics}.

\spara{Privacy issues.}\label{subsec:privacy-issues}
Social media, since it is, by definition, content directly created by end users, might carry  personally identifiable information (PII).  Researchers and practitioners should be mindful of any explicit or inferred PII in the data. For example, an individual may not explicitly reveal PII such as location for privacy reasons; inferring this user's location may be in conflict with her expectation of privacy.  As a result, academic work may have to be approved by a human subjects review process (ethics approval).
Practitioners, on the other hand, may be subject to a response organization's own standards and policies (see e.g.~\citeN{gilman:ocha-note} and \citeN{icrc:professional-standards}).

\section{Event Detection and Tracking}\label{sec:event-detection}

Most systems for social media processing during crises start with \textit{event detection}.
An event is the occurrence of something significant which is associated with a specific time and location~\cite{brants2003system}. However, due to the online nature of social media communications, events as they play out in social media may or may not be necessarily associated with a physical location.
In the context of social media, Dou et al.~\citeyear{dou2012leadline} define an event as: ``An occurrence causing changes in the volume of text data that discusses the associated topic at a specific time. This occurrence is characterized by topic and time, and often associated with entities such as people and location''.

Crisis and emergency situations typically fall into two broad categories: \emph{predicted} (or \emph{forewarned}) and \emph{unexpected}.
Some disaster events can be predicted to a certain level of accuracy based on meteorological or other data (e.g. this is the case with most large storms and tornadoes), and information about them is usually broadcast publicly before the actual event happens.  Sometimes an event may not be explicitly anticipated, for example as with a mass protest, but still may be forecast from social media and other data \cite{ramakrishnan:EMBERS}.
Other events cannot be predicted (e.g. earthquakes), and in this case an automatic detection method is useful to find out about them as quickly as possible once they happen.
In this section, we study techniques available for the automatic detection of both predicted, and unexpected events.

Historically, methods for event detection and tracking in social media in the context of crises and emergencies, are adaptations of methods to perform these tasks with a more general scope: that of detecting news. These methods, in turn, are adaptations of methods to find new topics in general document collections. 

%

\subsection{Background on Event Detection and Discovery}

A well-studied problem in Information Retrieval is detecting events in a stream of documents (see e.g. \citeN{allan2002topic}).
These documents can be news articles from traditional media sources, or posts on social media (e.g., tweets, Facebook posts, Flickr images).
Traditionally, the {\em Topic Detection and Tracking} (TDT) research community uses newswires as source data streams for event detection. 

Various techniques are employed in TDT including story segmentation, topic detection, new event detection, link detection, and topic tracking.
\emph{Story segmentation} focuses on determining story boundaries from streaming speech recognition output, usually from radio or television broadcasts.
\emph{Topic detection} groups related documents together into cohesive topics.
\emph{New event detection} processes each new document to decide if it describes a previously unseen story.
\emph{Link detection} detects that if two documents are similar or not.
Finally, \emph{event tracking} follows the evolution of an event/topic to describe how it unfolds. 

Event detection on social media is different from the traditional event detection approaches that are suitable for other document streams. Social media data emerge more quickly, and in larger volumes than traditional document streams. In addition, social media data are composed of short, noisy, and unstructured content that often require a different approach than what is used with traditional news articles.
Considering the unique characteristics of social media streams, we focus the remainder of this section on new event detection and event tracking.
Nevertheless, techniques and evaluation metrics from the TDT community provide insight into methods that might work for the Twitter domain.

\subsection{New Event Detection (NED)}

In the context of mass emergencies, New Event Detection (NED) refers to the task of discovering the first message related to an event by continuously monitoring a stream of messages.
%
%
NED decides whether a message is about something new that has not been reported in previous messages, or not~\cite{yang2009discovering}.
``New'' is normally operationalized as sufficiently different according to a similarity metric. Hellinger similarity, Kullback-Leibler divergence, and cosine similarity are among the metrics commonly used in NED~\cite{kumaran2004classification}.

As observed in Section~\ref{sec:retro-vs-online}, we can do retrospective (off-line) or on-line analysis~\cite{yang1998study}. While the most useful NED systems for emergencies are those who perform this analysis on-line, they are often adaptations or improvements of off-line methods, which we discuss next.

\subsubsection{Retrospective New Event Detection}

Retrospective NED refers to the process of identifying events using messages or documents that have arrived in the past. Methods for retrospective NED involve the creation of clusters of documents or messages based on a suitable definition of similarity between them, which may involve more than one dimension of them, e.g. using similar words, involving similar groups of people, occurring close to each other in time and/or space, etc.

For instance, 
\citeN{zhao2007temporal} introduce a retrospective NED method that uses textual, social and temporal characteristics of the documents to detect events on social streams such as weblogs, message boards, and mailing lists.
They build multi-graphs using social textual streams, where nodes represent social actors, and edges represent the flow of information between actors. Clustering techniques and graph analysis are combined to detect an event.

\citeN{sayyadi2009event} introduce a retrospective NED approach that overlays a graph over the documents, based on word co-occurrences. They assume that keywords co-occur between documents when there is some topical relationship between them. Next, a community detection method over the graph is used to detect and describe events.
\citeN{pohl2012automatic} describe a two-phase clustering approach to identify crisis-related sub-events in photo-hosting site Flickr and video-hosting site YouTube.
During the first phase, which is to identify sub-events, clusters are formed by using only items that contain geographical coordinates. These coordinates are added automatically by the device used to capture the photo or video, or are added later by its author/uploader. Next, they calculate term-based centroids of the identified clusters using cosine distance to further describe the identified sub-events.

Another retrospective NED approach is presented in~\citeN{chen2009event}, with experiments run on Flickr. It uses photos, user-defined tags, and other meta-data including time and location to detect events. This approach simultaneously analyzes the temporal and geographical distribution of tags, and determines the event type (e.g. whether it is recurring) to form clusters. Finally, for each tag cluster, the corresponding photos are retrieved.

%
\citeN{ritter2012open} extract significant events from Twitter by focusing on certain types of words and phrases. In their system, called {\em TwiCal}, they extract event phrases, named entities, and calendar dates. To extract named entities, they use a named entity tagger trained on 800 randomly selected tweets. To extract event mentions they use a Twitter-tuned part-of-speech tagger~\cite{ritter2011named}. The extracted events are classified retrospectively into event types using a latent variable model that first identifies event types using the given data, and then performs classification. 


Li et al.~\citeyear{li2012twevent} introduce \textit{Twevent}, a system that uses message segments instead of individual words to detect events. The authors claim that a tweet segment, which represents one or more consecutive words in tweets, contains more meaningful information than unigrams. The \textit{Twevent} approach works in phases. First, the individual tweet is segmented, and bursty segments are identified using the segments' frequency in a particular time window. Next, identified segments are retrospectively clustered using K-Nearest Neighbors (KNN) clustering. Finally, a post-filtering step uses Wikipedia concepts to filter the detected events.

\subsubsection{Online New Event Detection}

Online new event detection does not use previously seen messages or any prior knowledge about the events to be identified.
Online NED is typically performed with low latency (in real-time), in the sense that the time between seeing a document corresponding to a new event, and reporting that a new event has been detected, is relatively short.

\spara{Methods based on keyword burst.}
A straightforward approach is to assume that words that show sharp frequency increases over time are related to a new event.
For instance, Robinson et al.~\citeyear{robinson2013sensitive} introduce a system to detect earthquakes using Twitter. The earthquake detector, which is based on the Emergency Situation Awareness (\textit{ESA}) platform~\cite{power2014emergency}, checks for the keywords ``\texttt{earthquake}'' and ``\texttt{\#eqnz}'' in the real-time Twitter stream, and applies a burst detection method to analyze word frequencies in fixed-width time-windows and compare them to historical word frequencies. Unusual events are identified if the observed frequencies are much higher than those recorded in the past.
Earle et al.~\citeyear{earle2012twitter} have compared simple keyword-based approaches with data from seismological sensors, finding that while many earthquakes are not detected by Twitter users, detections are fast,  ``considerably faster than seismographic detections in poorly instrumented regions of the world.''


Marcus et al.~\citeyear{marcus2011twitinfo} introduced \textit{TwitInfo}, a system for detecting, summarizing and visualizing events on Twitter. \textit{TwitInfo} collects tweets based on a user-defined query (e.g. keywords used to filter the Twitter stream).
It then detects events by identifying sharp increases in the frequency of tweets that contain the particular user-defined query as compared to the historical weighted running average of tweets that contain that same query.
Further, tweets are obtained from the identified events to identify and represent an aggregated sentiment (i.e., classifying tweets into positive and negative classes).
The authors evaluated the system on various events such as earthquakes and popular football games.

A similar system, {\em TwitterMonitor}~\cite{mathioudakis2010twittermonitor}, also collects tweets from the Twitter stream and detects trends (e.g. emerging topics such as breaking news, or crises) in real-time. The trend detection approach proposed in their paper works in two phases. During the first phase, {\em TwitterMonitor} identifies bursty keywords which are then grouped  based on their co-occurrences. Once a trend is identified, additional information from the tweets is extracted to analyze and describe the trend. For example, the system uses Grapevine's entity extractor~\cite{angel2009s} which identifies entities mentioned in the trends.

Another Twitter-specific event detection approach introduced by~\citeN{petrovic2010streaming} uses Locality Sensitive Hashing (LSH) for hashing a fixed number of recent documents in a bounded space, and processed in a bounded time, to increase the performance of nearest neighbors search.  



With so many event detection systems, it is interesting to think about how they compare. McMinn et al.~\citeyear{McMinn:2013} describe a corpus to evaluate event detection methods, composed of 500 news events sampled over a four-week period, and including tweet-level relevance judgments for thousands of tweets referring to these events.  While existing NED systems have not been evaluated against this corpus, we anticipate this calibration of systems to occur in the future.  

\spara{Beyond keyword bursts.}
There are well-known problems of relying on increases in the frequency of a keyword (or a segment) to detect events.
For instance, consider popular hashtags such as ``\#musicmonday," which is used to suggest music on Mondays, or ``\#followfriday/\#ff,'' which are used to suggest people to follow on Fridays.
In these cases, there should be big pseudo-events detected every Monday and every Friday.

To address this problem, \citeN{becker2011beyond} present an approach to classify real-world events from non-events using Twitter. They use four types of features, which are temporal, social, topical, and Twitter-specific, to identify real events using the Twitter stream in real-time.
First, based on temporal features (i.e., volume of messages posted during an hour), they form initial clusters using the most frequent terms in the messages. Clusters are then refined using social features (i.e., users' interactions like re-tweets, replies, mentions). Next, they apply heuristics, for example, a high percentage of re-tweets and replies often indicates a non-event, whereas a high percentage of mentions indicates that there is an event.
Further, cluster coherence is estimated using a cosine similarity metric between messages and cluster centroid. 
Finally, as the authors report that multi-word hashtags (e.g. ``\#musicmonday'' and ``\#followfriday'') are highly indicative of some sort of Twitter specific discussion and do not represent any real event, they check the frequency of such hashtags used in each cluster to further refine the results. 

Weng et al.~\citeyear{weng2011event} present an algorithm for event detection from tweets using clustering of wavelet-based signals. Their approach involves three steps. First, they use wavelet transformation and auto correlation to find bursts in individual words, and keep only the words with high signal auto-correlations as event features. Then, the similarity for each pair of event-features is measured using cross correlation. Finally, they use a modularity-based graph partitioning algorithm to detect real-life events. One of the strong points of this approach over the traditional event detection approaches is the capability of differentiating real-life big events from trivial ones. This is achieved mainly by two factors: the number of words, and the cross-correlation among the words related to an event.

Unlike the approaches presented above, Corley et al.~\citeyear{corley2013social}  present a method to detect and investigate events through meta-data analytics and topic clustering on Twitter. 
Various features such as re-tweets, usage of different terms, and hashtags are analyzed for a certain time period to determine a baseline and a noise ratio. An event is detected once a particular feature value exceeds its noise boundaries and expected threshold.
%
Once an event has been detected, its related topics are identified using the topic clustering approach. 


\spara{Domain-specific approaches.}
As in many natural language processing applications, approaches that are specific to a certain domain generally perform better than the approaches that are open-domain or generic.

For instance, Phuvipadawat and Murata~\citeyear{phuvipadawat2010breaking} describe a method for detecting breaking news from Twitter. First, tweets containing the hashtag ``\#breakingnews'' or the phrase ``breaking news'' are fetched from the Twitter streaming API. Grouping of the extracted tweets is then performed, based on content similarity, and using a variant of the TF-IDF technique. Specifically, the similarity variant assigns a high similarity score to hashtags and proper nouns, which they identify using the Stanford Named Entity Recognizer (NER) implementation.

The authors consider three types of features associated with tweets: statistical features (i.e., number of words in a tweet, position of the query word within a tweet), keyword-based features (i.e., the actual words in a tweet), and contextual features (e.g., words appearing nearby the query term, for instance if ``earthquake'' is a query, terms such as ``magnitude'' and ``rocks'' would be features in the tweet ``{\em 5.3 magnitude earthquake rocks parts of Papua New Guinea}'').
In order to determine if a tweet corresponds to one of these hazards or crises, they use Support Vector Machines (SVM)---a known supervised classification algorithm (more about supervised classification in Section~\ref{sec:supervised_classification}).

Data from traditional media sources can also be used for detecting newsworthy events. Not surprisingly, traditional media and social media have different editorial styles, perceived levels of credibility, and response delays to events.
\citeN{tanev2012enhancing} find news articles describing security-related events (such as gun fights), and use keywords in their title and first paragraph to create a query. This query is issued against Twitter to obtain tweets related to the event.
\citeN{dou2012leadline} describe {\em LeadLine}, an interactive visual analysis system for event identification and exploration. 
{\em LeadLine} automatically identifies meaningful events in social media and news data using burst detection. Further, named entities and geo-locations are extracted from the filtered data to visualize them on a map through their interface.


Another domain-specific event-detection method is based on pre-specified rules and introduced in~\citeN{li2012tedas}. Their system, {\em TEDAS}, detects, analyzes, and identifies relevant crime and disaster related events on Twitter. First, tweets are collected based on iteratively refined rules (e.g., keywords, hashtags) from Twitter's streaming API. Next, tweets are classified via supervised learning based on content as well as Twitter-specific features (i.e., URLs, hashtags, mentions). Additionally, location information is extracted using both explicit geographical coordinates and implicit geographical references in the content. Finally, tweets are ranked  according to their estimated level of importance.

\citeN{sakaki2010earthquake} detect hazards and crises such as earthquakes, typhoons, and large traffic jams using temporal and spatial information. \emph{LITMUS}~\cite{musaevlitmus} detects landslides using data collected from multiple sources. The system, which depends on the USGS seismic activity feed provider, the TRMM (NASA) rainfall feed, and social sensors (e.g. Twitter, YouTube, Instagram), detects landslides in real-time by integrating multi-sourced data using relevance ranking strategy (Bayesian model). Social media data is processed in a series of filtering steps (keyword-based filtering, removing stop-words, geotagging, supervised classification) and mapped based on geo-information either obtained from meta-data or from content. Events with high relevancy are identified as ``real events.''

\subsection{Event Tracking and Sub-Event Detection}\label{subsec:event-tracking-subevent-detection}

\spara{Event tracking.}
Event tracking refers to the task of studying how events evolve and unfold. For a general discussion on the subject, see~\citeN{allan2002topic} and \citeN{lee2013event}.

The way in which emergency response agencies deal with crisis events varies as a crisis unfolds.
%
Emergency situations typically consist of four phases: warning, impact, emergency, and recovery~\cite[page 51]{killian_2002_introduction}. During the warning phase, the focus is on monitoring the situation. Impact is when the disaster agent is actually at work, while the emergency phase is the immediate post-impact period during which rescue and other emergency activities take place. Recovery is the period in which longer-term activities such as reconstruction and returning to a ``normal'' state occur.

Various techniques have been proposed to identify event phases. For instance, Iyengar et al.~\citeyear{iyengar2011content} introduce an approach to automatically determine different phases of an event on Twitter. The approach, which is mainly based on content-based features of tweets, uses an SVM classifier and a hidden Markov model. Various content-specific features such as bag of words, POS (part-of-speech) tags, etc. are used to automatically classify tweets into three phases of an event: {\em before}, {\em during}, and {\em after.} A disaster-specific lexicon of discriminative words for each phase of the event can also be employed~\cite{chowdhurytweet4act}.


\spara{Sub-event detection.}
The detection of large-scale events has been studied with more attention than the detection of small-scale ``sub-events'' that happen as a crisis unfolds.

\citeN{pohl2012automatic} show the importance of sub-event detection during crisis situations. They use multimedia meta-data (tags and title) associated with content found on social media platforms such as YouTube and Flickr. Their framework uses a clustering approach based on self-organizing maps to detect sub-events.
First, a pre-selection of the data is performed based on user-identified keywords. The selected data is then passed to a sub-event detection module that performs clustering to further split the data into sub-events.

\citeN{khurdiya2012extraction} present a system for event and sub-event detection using Conditional Random Fields (CRF)~\cite{lafferty2001conditional}. Their system consists of four main modules: (1) a CRF-based event extractor to first extract actor, action, date, and location; event titles are also extracted using using CRFs; (2) an event resolution to find similar events; (3) an event compiler that characterizes events; and (4) an event reporting module which is the end-user interface used to browse events details. 

\citeN{hua2013sted} introduce {\em STED}, a semi-supervised targeted-interest event and sub-event detection system for Twitter. To minimize the human effort required for labeling, they introduce an automatic label creation and expansion technique, which takes labels obtained from newspaper data and transfers them to tweets. They also propagate labels using mentions, hashtags, and re-tweets. Next, they build mini-clusters using a graph partitioning method to group words related to the event, and use supervised learning to classify other tweets using the examples provided by each mini-cluster. A final step on the classified output is to perform a location estimation using information from geo-coded tweets.

\begin{table}
\caption{Some of the event detection tools surveyed. The table includes the types of events for which the tool is built (open domain or specific), whether detection is performed in real-time, the type of query (open or ``kw''=keyword-based), and whether it has spatio/temporal or sub-event detection capabilities. Sorted by publication year.}
\label{tbl:event_detection_tools}
\centering\scriptsize
\begin{tabular}{  p{1.4cm} p{2.5cm} p{1.7cm} p{0.2cm} p{0.3cm} p{0.7cm} p{0.3cm} p{3.8cm} }
\toprule
System/tool & Approach & Event types & Real-time & Query type & Spatio-temporal & Sub-events & Reference \\
\midrule
\textit{TwitterMonitor} & burst \mbox{detection} & open domain & yes & open & no & no & \cite{mathioudakis2010twittermonitor} \\
\textit{TwitInfo} & burst \mbox{detection} & earthquakes+ & yes & kw & spatial & yes & \cite{marcus2011twitinfo} \\
\textit{Twevent} & \mbox{burst segment detection} & open domain & yes & open & no & no & \cite{li2012twevent} \\
\textit{TEDAS} & \mbox{supervised classification} & crime/disasters & no & kw & yes & no & \cite{li2012tedas} \\
\textit{LeadLine} & \mbox{burst detection} & open domain & no & kw & yes & no & \cite{dou2012leadline} \\
\textit{TwiCal} & \mbox{supervised classification} & \mbox{conflicts/politics} & no & open & temporal & no & \cite{ritter2012open} \\
\textit{Tweet4act} & \mbox{dictionaries} & disasters & yes & kw & no & no & \cite{chowdhurytweet4act} \\
\textit{ESA} & burst \mbox{detection} & open domain & yes & kw & spatial & no & \mbox{\cite{robinson2013sensitive}} \\
\bottomrule
\end{tabular}
\end{table}

Table~\ref{tbl:event_detection_tools} lists event detection systems. The majority of them are surveyed above; additional tools are covered in the following sections.

\subsection{Challenges}
In addition to the discussion of general data challenges in Section~\ref{sec:data-challenges}, the following are particularly relevant to event detection.

\spara{Inadequate spatial information.}
Spatial and temporal information are two integral components of an event. Most systems that rely on Twitter data for event detection face challenges to determine geographical information of tweets that lack GPS information. In this case, automatic text-based geotagging can be used, as described on Section~\ref{sec:geotag}.

\spara{Mundane events.}
People post mundane events on social media sites. These data points introduce noise, which creates further challenges for an event detection algorithm to overcome. For such cases, separation of real-life big events from trivial ones is required.

\spara{Describing the events.}
Creating a description or label for a detected event is in general a difficult task. Often the keywords that are more frequent during the event are presented as a description for the event in the form of a list of words (e.g. \{ \emph{sandy}, \emph{hurricane}, \emph{new york} \} ), but that list does not constitute a grammatically well-formed description (e.g. ``\emph{Hurricane Sandy hits New York}'').  We will see one approach to address this in Section \ref{sec:summarization}.

\section{Clustering, Classification, Extraction and Summarization}
\label{sec:finding_information}

Once we have found social messages related to a crisis event or topic, there are several ways in which we might process them.  In this section, we describe some approaches.  Broadly, we separate techniques into those classifying the data item as a whole and those extracting useful information  from the content of one or more data items.  

\subsection{Classifying Social Media Items}
In many situations, we are interested in classifying social media items into one or more categories.  In this section, we will describe three methods of classification with decreasing levels of manual human supervision.  Text classification is a large field of research, so we will cover literature relevant to disasters.  Interested readers should refer to other sources for a broader discussion \cite{Srivastava:2009:TMC:1571651}.

\subsubsection{Content Categories}\label{subsec:content-categories}

There is not a single standard or widely-accepted way of categorizing crisis-related social media messages. While some crisis-related ontologies have been proposed (see Section~\ref{sec:ontologies}), in general, different works use different approaches.

Table~\ref{tbl:event_dimensions_summary} summarizes various dimensions of social media content that different research articles have used to classify information:
\begin{compactenum}
 \item By factual, subjective, or emotional content: to separate between facts (or combinations of facts and opinions), from opinions, or expressions of sympathy.
 \item By information provided: to extract particular categories of information that are useful for various purposes.
 \item By information source: to select messages coming from particular groups of users, e.g. eyewitness accounts or official government sources.
 \item By credibility: to filter out messages that are unlikely to be considered credible.
 \item By time: to filter messages that refer to different stages of an event, when temporal boundaries for the event are unclear.
 \item By location: to select messages according to whether they originate from or near the place that was affected by an event, or from areas that were not affected.
\end{compactenum}

Other classification dimensions can be envisioned. In general, we observe that the selection of the set of categories used by researchers and practitioners is usually driven by two main factors: the data that is present in social media during crises, and the information needs of response agencies.
None of these factors is static, as both can change substantially from one crisis situation to another.
We also remark that different types of disaster elicit different distributions of messages~\cite{olteanu_2015_what}.

\begin{table}[p]
\caption{Classification of various dimensions of content posted on social media during high impact events with description and related work references.}
\label{tbl:event_dimensions_summary}
\centering \scriptsize
\begin{tabular}{  p{3.0cm} p{9.8cm} }
\toprule

Classification dimension & Description/examples \\ 	   \midrule

\multicolumn{2}{l}{\bf By factual, subjective, or emotional content}  \\
Factual information		& \textit{(Examples under ``By information provided'')}\\
Opinions	& opinions, criticism (e.g. of government response) \\
Sympathy	& condolences, sympathy~\cite{kumar2013whom}; condolences~\cite{acar2011twitter}; support~\cite{Hughes:2014}; thanks, encouragement~\cite{bruns2014crisis}; prayers~\cite{olteanu2014crisislex} \\
Antipathy	& \textit{schadenfreude}, animosity against victims (e.g. because of long-standing conflict)\\
Jokes		&  jokes, trolling~\cite{metaxas2013rise}\\[4pt]

\multicolumn{2}{l}{{\bf By information provided}}   \\
Caution and advice & caution and advice~\cite{imran2013extracting}; warnings~\cite{acar2011twitter}; hazard, preparation~\cite{olteanu2014crisislex}; tips~\cite{leavitt2014upvoting}; advice~\cite{bruns2014crisis}; status, protocol~\cite{Hughes:2014} \\[2pt]

Affected people & people trapped, news~\cite{caragea2011classifying}; casualties, people missing, found or seen~\cite{imran2013extracting}; self reports~\cite{acar2011twitter}; injured, missing, killed~\cite{Vieweg:2010}; looking for missing people~\cite{qu2011microblogging} \\[2pt]

Infrastructure/utilities & infrastructure damage~\cite{imran2013extracting}; collapsed structure~\cite{caragea2011classifying}; built environment~\cite{Vieweg:2010}; closure and services~\cite{Hughes:2014} \\[2pt]

Needs and donations & donation of money, goods, services~\cite{imran2013extracting}; food/water shortage~\cite{caragea2011classifying}; donations or volunteering~\cite{olteanu2014crisislex}; help requests, relief coordination~\cite{qu2011microblogging}; relief, donations, resources~\cite{Hughes:2014}; help and fundraising~\cite{bruns2014crisis} \\[2pt]

Other useful information & hospital/clinic service, water sanitation~\cite{caragea2011classifying}; help requests, reports about environment~\cite{acar2011twitter}; consequences~\cite{olteanu2014crisislex} \\ \\

\multicolumn{2}{l}{\bf By information source}  \\
Eyewitnesses/Bystanders & members of public~\cite{metaxas2013rise}, victims, citizen reporters, eyewitnesses~\cite{diakopoulos2012finding,olteanu2014crisislex,bruns2014crisis} \\
Government & administration/government~\cite{olteanu2014crisislex}; police and fire services~\cite{Hughes:2014}; government~\cite{bruns2014crisis}; news organization and authorities~\cite{metaxas2013rise} \\
NGOs & non-government organizations~\cite{DeChoudhury:2012}\\
News Media & news organizations and authorities, blogs~\cite{metaxas2013rise}, journalists, media, bloggers~\cite{DeChoudhury:2012}; news organizations~\cite{olteanu2014crisislex}; professional news reports~\cite{leavitt2014upvoting}; media~\cite{bruns2014crisis} 
\\[4pt]

\multicolumn{2}{l}{\bf By credibility}  \\
Credible information & newsworthy topics, credibility~\cite{castillo2013predicting}; credible topics~\cite{canini2011finding}; content credibility~\cite{gupta2012credibility}; users and content credibility~\cite{gupta2014tweetcred}; source credibility~\cite{thomson2012trusting}; fake photos~\cite{gupta2013faking} \\
Rumors & rumor~\cite{Hughes:2014,castillo2013predicting}\\\\

\multicolumn{2}{l}{\bf By time}  \\
Pre-phase/preparedness & posted before an actual event occurs, helpful for the preparedness phase of emergency management~\cite{petak1985emergency}; pre-disaster, early information~\cite{iyengar2011content,chowdhurytweet4act}\\
Impact-phase/response & posted during the impact phase of an event, helpful for the response phase of emergency management~\cite{petak1985emergency}; during-disaster~\cite{iyengar2011content,chowdhurytweet4act}\\
Post-phase/recovery & posted after the impact of an event, helpful during the recovery phase~\cite{petak1985emergency}; post-disaster information~\cite{chowdhurytweet4act,iyengar2011content} \\ \\

\multicolumn{2}{l}{\bf By location}  \\
Ground-zero & information from ground zero (victims reports, bystanders)~\cite{longueville2009omg,ao2014estimating} \\
Near-by areas & information originating close to the affected areas~\cite{longueville2009omg} \\
Outsiders & information coming from other parts of world, sympathizers~\cite{kumar2013whom}; distant witness (in the sense of~\cite{carvin_2013_distant}); not on the ground~\cite{starbird2012learning}; location inference~\cite{ikawa2012location}\\

\bottomrule
\end{tabular}
\end{table}

\subsubsection{Supervised Classification} \label{sec:supervised_classification}

When a set of example items in each category is provided, a \emph{supervised classification} algorithm can be used for automatic classification.
This type of algorithm `learns' a predictive function or model from features of these  examples (see Section \ref{subsec:feature-extraction})  in order  to label new, unseen data items.   This set of examples is referred to as the \textit{training set}.  
After a model has been learned from the training data, it is evaluated using a different, hold-out set of labeled items, not used during the training.  This second set of examples is referred to as the \emph{testing set}.

%
Depending on the nature of data in hand, different pre-processing techniques can be used. 
In any case, the input items are transformed into feature vectors, following the methods described in Section~\ref{subsec:feature-extraction}.

\spara{Training examples.}
The number of training examples required to achieve good accuracy depends on many factors, including the number of categories into which messages have to be classified, and the variability of messages inside each category. Typical sizes of training sets range from a few hundred~\cite{yin2012using} to a few thousand~\cite{imran2014aidr}. More examples yield better results in general, with diminishing results after a certain point.
In general, the accuracy of models created using training data from one crisis decreases when applied to a different crisis, or when applied to the same crisis but at a different point in time~\cite{imran2014coordinating}.

\spara{Feature selection.}
Even if messages are brief, the feature space in which they are represented is typically high dimensional (e.g. one dimension for every possible term). This introduces a number of problems including the amount of computational resources required for the data analysis, and it also increases the chances of over-fitting the training data. In this case, a \emph{feature selection method} (e.g. mutual information) should be employed as a first step to discard input features that have little or no correlation with the given training labels.  Feature selection is an active area of machine learning research and state of the art techniques can be found in modern textbooks or journals (e.g.~\citeN{Guyon:JMLR-feature-selection-special-issue}).

\spara{Learning algorithms.}
After features have been extracted and selected, a machine learning algorithm can be applied. Supervised classification algorithms include, among others, na{\"i}ve Bayes, Support Vector Machines (SVM), logistic regression, decision trees, and random forests.
The choice of a method is largely dependent on the specific problem setting. For instance, \emph{ESA}~\cite{yin2012using,power2014emergency} uses na{\"i}ve Bayes and SVM, \emph{EMERSE}~\cite{caragea2011classifying} uses SVM, \emph{AIDR}~\cite{imran2014aidr} uses random forests, and \emph{Tweedr}~\cite{ashktorab2014tweedr} uses logistic regression.
While in most cases algorithms are used to predict a single label for each element, adaptations of these algorithms that generate multiple labels for each element are sometimes employed (e.g.~\citeN{caragea2011classifying}).

\spara{Ensemble/stacked classification.}
In some cases an explicit model of a certain factor is desired, as exemplified by the work of \citeN{verma2011natural}. They observe that messages that contribute the most to situational awareness are also those that are expressed using objective (as opposed to subjective) language. In this case, one can create a \emph{stacked classifier} in which at one level certain characteristics of the message are modeled (e.g. by having a classifier that classifies messages as objective or subjective) and at the next level these characteristics are combined with other characteristics also modeled by specific classifiers (e.g. writing styles such as formal or informal), and with features from the message itself. \citeN{verma2011natural} find that this approach performs better than directly using the input features.

Comparison in terms of classification effectiveness/accuracy is a non-trivial task. This is due to the use of different datasets, different baselines, and different performance measures (e.g. AUC, F1, precision). However, classification accuracies reported by most of the systems we have reviewed for this survey range from 0.60 to 0.90.
Recent collections available for research may contribute to make systems easier to compare.\footnote{\url{http://crisislex.org/}}

\subsubsection{Unsupervised Classification} \label{sec:unsupervised_classification}

Clustering is an \emph{unsupervised machine learning method}; a family of methods that seek to identify and explain important hidden patterns in unlabeled data. Unsupervised machine learning methods include clustering, dimensionality reduction (e.g. principal component analysis), and hidden Markov models, among others.

The process of performing clustering, 
begins by ingesting a set of items (e.g., documents, tweets, images) which are then processed with the objective of grouping similar items together.
In general, the goal is to form clusters in such a way that elements within a cluster are more similar to each other than to the elements that belong to other clusters. Many clustering algorithms have been developed based on different approaches, examples include \emph{K-means} (centroid-based), \emph{hierarchical clustering} (connectivity-based), \emph{DBSCAN} (density-based), among many others. For an overview of clustering methods, see~\citeN[Part III]{zaki_meira_2014_datamin}.

In the context of dealing with social media data during crises, clustering can help reduce the number of social media messages that need to be processed/examined by humans, for instance by displaying multiple equivalent messages as a single item instead of multiple ones.
This is the approach used by \emph{CrisisTracker}~\cite{rogstadius2013crisistracker}, which is a crowdsourced social media curation system for disaster awareness. The system, which collects data from Twitter based on predefined filters (i.e., keywords, bounding box), groups these tweets into \emph{stories}, which are clusters of tweets. These stories are then curated/classified by humans, whose effort can be greatly reduced if they are asked to classify entire clusters instead of single tweets.
The specific clustering method employed in this case is locality-sensitive hashing, an efficient probabilistic technique that uses hash functions to detect near-duplicates in data. 
Another example is \textit{SaferCity}~\cite{berlingeriosafercity}, which identifies and analyzes incidents related to public safety in Twitter, adopting a spatio-temporal clustering approach based on the modularity maximization method presented in~\citeN{blondel2008fast} for event identification. Clusters are then classified using a semantic labeling approach using a controlled vocabulary 
and based on a rank score provided by the Lucene library.\footnote{http://lucene.apache.org/}

In addition to clustering methods that partition the items into groups, there are \emph{soft clustering} methods that  allow an item to simultaneously belong to several clusters with varying degrees.  This is the case of topic modeling methods, out of which Latent Dirichlet Allocation (LDA) is one of the most popular. In the crisis domain, \citeN{kireyev2009applications} use topics extracted using LDA, including a weighting scheme that accounts for the document frequency (number of tweets containing a word) of the words in the tweet, as well as for the length of the tweet. When applied to data from an Earthquake in Indonesia in 2009, it detects topics that cover different aspects of the crisis such as \{ \emph{tsunami}, \emph{disaster}, \emph{relief}, \emph{earthquake} \}, \{ \emph{dead}, \emph{bodies}, \emph{missing}, \emph{victims} \}
 and \{ \emph{aid}, \emph{help}, \emph{money}, \emph{relief} \}.
There is free software available for creating these topics models, such as MALLET~\cite{mccallum2002mallet}, which has been applied to crisis data by~\citeN{karandikar2010clustering}.

\subsubsection{Discussion}
In the context of social media data analysis, many factors influence the choice of a learning approach (i.e. supervised vs. unsupervised). For instance, approaches based on unsupervised learning certainly have advantages over supervised ones in cases where obtaining training examples would be prohibitively costly, would introduce an unacceptable delay or would simply not be possible. Unsupervised approaches are more useful in cases where information classes are completely unknown from a information seeker point of view. 

However, in terms of usefulness, the output of an unsupervised approach might be considered less useful than an approach based on supervised learning. Emergency responders may be used to specific categorizations and methodologies (e.g. the MIRA framework used by the United Nations\footnote{\url{https://docs.unocha.org/sites/dms/Documents/mira_final_version2012.pdf} (accessed Nov. 2014).}) and may expect that software tools output the same categories. 

\subsection{Sub-Document Analysis}
In contrast to text classification schemes which make predictions about data items, sub-document analysis techniques extract granular information from the \textit{content} of the data items.  In this section we present two important sub-document analysis methods in the context of crisis situations: information extraction and text summarization.

\subsubsection{Information Extraction}\label{subsec:information_extraction}

The task of automatically extracting structured information from unstructured (e.g., plain text documents) or semi-structured (e.g., web pages) data is known as Information Extraction (IE).
The most common information extraction task is named entity extraction, which consists of detecting regions of a text referring to people, organizations, or locations~\cite{liu2011recognizing,ritter2011named}. This is the first step towards semantic enrichment (see Section~\ref{subsec:semantic-enrichment}).

In the context of crisis-related social media, information extraction can be used, for instance, to transform tweets reporting injured people in natural language (e.g. ``\emph{5 injured and 10 dead in Antofagasta}'') to normalized records such as \{$<$people-affected=5, report-type=injury, location=Antofagasta, Chile$>$, $<$people-affected=10, report-type=fatal-casualty, location=Antofagasta, Chile$>$\}. 
These records are machine-readable, which means they can be easily filtered, sorted, or aggregated.

Information extraction from social media is a challenging task because of the informal writing style and the presence of many ungrammatical sentences, as was noted in Section~\ref{subsec:content-issues}.
State-of-the-art approaches to information extraction involve the use of probabilistic sequential models such as hidden Markov models, conditional Markov models, maximum-entropy Markov models, or conditional random fields. Heuristics based on regular expressions can also be applied to this problem, although these are in general less effective than probabilistic methods.

Varga et al.~\citeyear{varga2013aid} use linguistic patterns and supervised learning to find ``trouble expressions'' in social media messages. For instance, in a tweet such as ``\emph{My friend said infant formula is sold out. If somebody knows shops in Sendai-city where they still have it in stock, please let us know},'' the nucleus of the problem is the sentence ``\emph{infant formula is sold out}.'' This extraction is then used to match tweets describing problems to tweets describing solutions to those problems.

Imran et al.~\citeyear{imran2013practical,imran2013extracting} apply conditional random fields to the extraction of information from tweets. Their method proceeds in two steps. First, tweets are classified to consider categories such as ``infrastructure damage,'' ``donations,'' and ``caution and advice.''
Then, a category-dependent extraction is done, where for instance for ``infrastructure damage'' tweets, the specific infrastructure reported damaged is extracted, while for ``donations'' tweets, the item being offered in donation is extracted.

\subsubsection{Summarization}\label{sec:summarization}
Another approach to dealing with information overload involves presenting users with a text-based representation of the evolving event.  Text summarization  refers to the generation of a text summary of a document or set of documents \cite{nenkova:summarization-survey}. Text summarization systems are optimized to generate a text summary that contain only the core topics discussed in the set of documents. Most systems produce this summary by \textit{extracting} key sentences from the input document.  This is in contrast to systems that produce a summary by  \textit{abstracting} or generating new sentences.

During crisis events, text summarization must be done in an \textit{incremental} and \textit{temporal} manner.
Incremental text summarization, also known as \textit{update summarization}, refers to generating a summary given: 
\begin{inparaenum}
\item the set of documents to be summarized, and
\item a reference set of documents which the user has read.
\end{inparaenum}
The objective for the system is to produce a summary only of data the user has not already read~\cite{tac2008:update-summarization-overview}. Temporal text summarization refers to creating an extractive summary from a set of time-stamped documents, usually in retrospect~\cite{allan:temporal-summaries,nallapati:event-threading,feng:cikm2009}.

Drawing on the work from summarization research, the TREC Temporal Summarization track focuses on generating updates relating to unfolding crisis events immediately after their occurrence~\cite{trec2013:temporal-summarization-overview}. Table~\ref{tab:trec-ts-example} presents an example of this type of summary. The focus of this initiative is to first define standardized metrics for the task and then to encourage the development of systems optimizing them.
These metrics include time-sensitive versions of precision and recall, ensuring that systems are penalized for \textit{latency}: delivering information about an event long after it occurred.  In addition, the metrics include a redundancy penalty to prevent systems from delivering repetitive information. In order to optimize these metrics, systems can use staged text analysis with standard information retrieval measures \cite{trec2013:terrier}.
Alternatively, systems can use regression-based combinations of features from classic text summarization literature \cite{qi:temporal-summarization,trec2013:hltcoe}. 
Other methods are purely content-based, hierarchically clustering sentence text~\cite{wang:update-summarization}. In the context of social media, Shou et al.~\citeyear{Shou:sumblr} propose a system for online update summarization based on incremental clustering.  The performance is evaluated under experimental conditions different from the TREC track, making it difficult to compare with other results. 

Although algorithms for summarization exist for crisis events, their development is still preliminary and several challenges remain. Foremost, the relative importance of different features is not well understood.  To date, research has primarily used features from batch text summarization.  A second challenge is scale.  Many algorithms require aggressive inter-sentence similarity computation, a procedure which scales poorly.  
\begin{table}[t]
\caption{Example summary from the TREC 2013 Temporal Summarization Track. Updates reflect new or updated information as it is reported.  }
\label{tab:trec-ts-example}
\centering
{\scriptsize
\begin{tabular}{p{0.9in}p{4.2in}}
\toprule
time & update \\
\midrule
Nov 21, 2012 10:52        &       Tel Aviv bus bombing; 13 injuries; reported on bus line 142; occured on Shaul Hamelech street; No claims of responsibility; 3 badly hurt; occured in the heart of Tel Aviv near military hdqtrs \\
Nov 22, 2012 20:49        &       occurred in an area with many office buildings; occured in area with heavy pedestrian traffic; first notable bombing in Tel Aviv since 2006; At least 28 people were wounded; Hamas' television featured people praising the attack; Khaled Mashal, leader of Hamas, categorically rejected any connection of the bombing to his group; ... \\ 
Nov 26, 2012 04:33        &       an Israeli Arab man was arrested on charges of planting the explosive device on the bus; Suspect was reportedly connected to Hamas; Suspect was reportedly connected to the Islamic Jihad \\
Nov 26, 2012 14:49        &       The Romanian Foreign Minister condemned the bombing,    \\
Nov 29, 2012 04:55        &       govt rep refers to attack as terrorist attack   \\
Nov 30, 2012 05:22        &       Fears about Bus Bomb Before the Cease-Fire: Could derail peace talks    \\
Nov 30, 2012 06:47        &       The suspect remotely detonated the explosive device; suspect hid device  in advance on the bus; The explosive device contained a large quantity of metal shrapnel designed to cause maximum casualties; The suspect later on confessed to carrying out the bus attack; suspect prepared the explosive device; suspect chose the target of the attack; ... \\
\bottomrule
\end{tabular}}
\end{table}

\subsection{Challenges}\label{sec:datamin-challenges}

In addition to the discussion on generic data-related challenges in Section~\ref{sec:data-challenges}, the following are particularly relevant to mining: 

\spara{Combining manual and automatic labeling.}
In a supervised learning setting, human labels are necessary, but they may be costly to obtain. This is particularly problematic in crises that attract a multilingual population, or for tasks that require domain knowledge (e.g. people who know informal, local place names in Haiti, and who speak Haitian Creole).
Labeled data are not always reliable, and may not be available at the time of the disaster; in this case, a hybrid approach that mixes human labeling and automatic labeling can be employed~\cite{imran_2013_engineering}. The selection of items to be labeled by humans can be done using \emph{active learning}, a series of methods to maximize the improvement in classification accuracy as new labels are received.


\spara{Domain adaptation.}
Ideally, one would like to avoid having to re-train an automatic classifier every time a new crisis occurs. However, simply re-using an existing classifier trained on data from a previous crisis does not perform well in practice, as it yields a substantial loss in accuracy, even when the two crises have several elements in common~\cite{imran2013practical}. 

In machine learning, \emph{domain adaptation} (or \emph{domain transfer}) is a series of methods designed to maximize the accuracy of a classifier trained on one dataset, adapting it to continue to perform well on a dataset with different characteristics. To the best of our knowledge, these methods have not been applied to crisis-related social media data.

\section{Semantic Technologies in Disaster Response}\label{sec:semantics}

One of the main goals of semantic technologies is to allow users to easily search through complex information spaces, and to find, navigate, and combine information. In the context of social media use during crises and mass convergence events, this is achieved by \emph{linking} data elements to concepts in a machine-readable way, enabling the representation of a situation as a complex and interrelated set of elements.

\subsection{Semantic Enrichment of Social Media Content}\label{subsec:semantic-enrichment}

Semantic technologies are particularly useful in social media, because they provide a powerful method for dealing with the variety of expressions that can be used to refer to the same concept, and with the many relationships that can exist between concepts.
For instance, suppose we are looking for messages related to infrastructure damage using a keyword-query search. We would think that searching for something such as ``damage AND (airport OR port OR bridge OR building ...)'' would be sufficient, until we notice that it is not only difficult to cover every particular type of infrastructure (airport, port, bridge, building, etc.) but also to cover every particular instance of that type (for instance, there are tens of thousands of airports in the world).

\emph{Named entity linking} is a widely-used semantic technology that deals with the above problem. It operates in two phases. First, a \emph{named entity recognizer} module detects entities---such as names of persons, places, and organizations. Second, for each named entity that is found, a \emph{concept} is located that more closely matches the meaning of that named entity in that context.
Concepts are generally operationalized as Uniform Resource Locators (URLs). For instance, \citeN{zhou2010resolving} link named entities to the URLs of articles on Wikipedia. In a phrase such as ``\emph{Terminal 2 of JFK was damaged},'' the named entity corresponding to the segment ``JFK'' would be linked to the URL \url{https://en.wikipedia.org/wiki/John_F._Kennedy_International_Airport}.
There are several free, and commercial services that can be used to perform named entity linking, including
Alchemy,\footnote{\url{http://www.alchemyapi.com/}}
Open Calais,\footnote{\url{http://www.opencalais.com/}} and
Zemanta.\footnote{\url{http://developer.zemanta.com/}}

Once an element is linked to a concept, further automatic annotation can be done by following links from the concept. Returning to our example of ``\emph{Terminal 2 of JFK was damaged},'' if we go to its Wikipedia page,\footnote{Or to its related semantic resource, DBPedia \url{http://dbpedia.org/}.} we can learn that ``JFK'' is an instance of the class ``airport.'' The airport concept is represented in this case by the URL of a Wikipedia category page to which the JFK Airport page belongs (\url{https://en.wikipedia.org/wiki/Category:Airports}), which in turn belongs to the category of transport building and structures (\url{https://en.wikipedia.org/wiki/Category:Transport_buildings_and_structures}).

After named entity linking, messages that have been semantically enriched can be used to provide \emph{faceted search}, a popular approach to interactively search through complex information spaces. In faceted search the information of interest can be found not only by specifying a related keyword, but also by specifying a concept or concepts associated with the  items of interest. In the example, we could select ``airport'' from a list of buildings and structures and then find a series of social media messages that are relevant, but that do not necessarily include the specific word ``airport'' in them.

\citeN{abel2011leveraging}, present an adaptive faceted search framework to find relevant messages on Twitter. The framework enriches tweets with semantics by extracting entities (i.e. persons, locations, organization), and then finding and linking those entities with external resources to create facets. Each facet enables search and navigation of relevant semantically related content.
In follow-up work~\cite{abel2012semantics}, they introduce \emph{Twitcident}, a system that supports semantic filtering, faceted search and summarization of tweets.
The semantic-based approach is also implemented in \emph{Twitris 2.0}~\cite{jadhav2010twitris} and \emph{EDIT}~\cite{traverso_2014_edit}, which present event-related social media capturing semantics in terms of spatial, temporal, and thematic dimensions. 

\subsection{Ontologies for Disaster Management}\label{sec:ontologies}

Information technologies to support disaster response often involve interactions between software operated by different agencies, and/or provided by different developers or vendors.
Allowing computer systems to communicate information in a unified way is a key challenge in general, but especially during crisis events where different agencies must address different dimensions of a problem in coordination with each other~\cite{hiltz2011introduction}.
Interoperability at the semantic level requires centralized specifications describing machine-understandable common vocabularies of concepts and linkages between them. An effective way of achieving this is to use machine-understandable ontologies that define, categorize and maintain relationships between different concepts to facilitate common understanding, and unified communication.

\begin{table}
\caption{Crisis ontologies, including some of the classes and attributes they cover, and the format in which they are specified (OWL: Ontology Web Language, RDF: Resource Description Framework).}
\label{tbl:crisis_ontologies}
\centering \scriptsize
\begin{tabular}{  p{1.2cm}  p{6.2cm} p{.5cm} p{1.7in}}
    \toprule
Ontology Name & Coverage & Format & Reference \\ \midrule
SOKNOS & resources, damage, disasters & OWL & \cite{babitski2011soknos} \\
HXL & damage, geography, organization, disasters & RDF & \url{http://hxl.humanitarianresponse.info/} \\
SIADEX & processes, resources, geography & RDF & \cite{de2005siadex} \\
OTN & specific to infrastructure & OWL & \cite{lorenz2005ontology} \\
MOAC & damage, disasters, processes, resources & RDF & \url{http://observedchange.com/moac/ns/} \\
FOAF & emergency management people & RDF & \url{http://www.foaf-project.org/} \\
AktiveSA & \mbox{transportation, meteorology, processes, resources, people} & OWL & \cite{smart2007aktivesa} \\
IntelLEO & response organizations & RDF & \url{http://www.intelleo.eu/} \\
ISyCri & damages, processes, disasters & OWL & \cite{benaben2008metamodel} \\
WB-OS & features, components and information to build crisis management web sites & XML & \cite{chou2011ontology}\\
EDXL-RM & \mbox{data exchange language for resource management} & XML & \url{https://www.oasis-open.org/} \\
\bottomrule
\end{tabular}
\end{table}

Table~\ref{tbl:crisis_ontologies} lists some of the ontologies that have been introduced in recent years.
%
Some examples from this table:

\begin{itemize}

\item The {\em Humanitarian eXchange Language} (HXL)\footnote{\url{http://hxl.humanitarianresponse.info/}} is an ontology created in 2011 and 2012 and is currently under review; it describes 49 classes and 37 properties. The focus of HXL is mainly on four areas: organization (i.e., formal response agencies like military, charities, NGOs), disaster (i.e., classification of disasters such as natural, man-made), geography (i.e., event location, geo-location of displaced people), and damage (i.e., damages related to humans, infrastructure).

\item \emph{Management of A Crisis} (MOAC)\footnote{\url{http://observedchange.com/moac/ns/}} is an ontology with 92 classes and 21 properties covering four areas: disaster, damage, processes (i.e., rescue, search, evacuation processes), and resources (i.e., services, vehicles, tents). HXL and MOAC have elements in common: in both cases the objective is to describe different aspects of a crisis, including its effects, the needs of those affected, and the response to the crisis. 

\item \emph{Integrated Data for Events Analysis} (IDEA)\footnote{\url{http://vranet.com/IDEA.aspx}} is a framework for coding social, economic and political events. It is used in the {\em Global Database of Events, Language and Tone} (GDELT),\footnote{\url{http://gdeltproject.org/}} which is a machine-generated list of event data extracted from news reports.

\item \emph{Service-orientierte ArchiteKturen zur Unterst{\"u}tzung von Netzwerken im Rahmen Oeffentlicher Sicherheit} (``Service-oriented architectures supporting networks of public security,'' SOKNOS)\footnote{http://soknos.de/} is an ontology for information integration for resource planning, including damage and resource categorization during disasters.


\end{itemize}

To support communication among different ontology-based systems, the problem of ontology heterogeneity needs to be solved by performing an \emph{ontology mapping}, which is the process of mapping the concepts of two ontologies from the same or from overlapping domains.
Many approaches have been proposed to perform ontology mapping. For instance,~\citeN{tang2006using} treated this as a decision-making problem and proposed an approach based on Bayesian decision theory. For a survey on ontology mapping techniques, see~\citeN{noy2004semantic}.

The ontologies in Table~\ref{tbl:crisis_ontologies} are crisis-specific, but not social-media specific. However, they can be combined with ontologies describing social media concepts such as users, tagging, sharing, and linking.
For instance, the ``Semantically Interlinked Online Communities'' (SIOC)\footnote{http://sioc-project.org/} ontology, originally developed to model sites such as blogs and online forums, has recently been extended to support the modeling of microblogs by adding concepts such as \emph{follower} or \emph{follows}.
An ontology specific to Twitter appears in~\citeN{celino2011making} and includes user sentiments and locations, while~\citeN{passant2008meaning} enables semantic tagging of social media data through an ontology called Meaning-Of-A-Tag (MOAT).
For a survey on ontologies developed for social media, see~\citeN{bontcheva2012making}.

\section{Summary and Future Research Directions}\label{sec:conclusions}

In a relatively short time period---roughly 4 to 6 years---the research community working on the topics we have covered here has achieved a fairly high degree of maturity with respect to filtering, classifying, processing, and aggregating social media data during crises.

However, the underlying (although sometimes explicitly stated) claim behind this line of work, i.e. that this research is \emph{useful} for the public and/or formal response agencies, that it has the potential to save lives and/or property during an emergency, remains to be seen.
%
While there are notable exceptions including the American Red Cross, the US Federal Emergency Management Agency, UN Office for the Coordination of Humanitarian Affairs and the Filipino Government, among others, the use of social media is still experimental for many organizations and not yet part of their normal, day-to-day operations.

There are two main directions in which we see future research going. First, continue deepening the data processing capabilities that have been the main focus of computing research on this topic thus far. Second, engage more deeply with human-centered approaches toward making the computing research the foundation of viable systems that emergency responders can implement. 

\subsection{Deepening Data Processing Capabilities}

\spara{From situational awareness to decision support.}
The systems we described in Section~\ref{sec:systems} for processing social media during disasters have a strong focus on situational awareness, which is an important first step but might not be enough for emergency management.
During an emergency, social media is used as an information source in order to make decisions. Therefore, next-generation systems should be designed and evaluated in terms of their decision-support capabilities. This might even include forecasting using signals from social media.

\spara{Extending to other types of media.}
Data from various sources should be processed and integrated: ``The strategies of emergency services organizations must also recognize the significant interweaving of social and other online media with conventional broadcast and print media.''~\cite{bruns2014crisis}. 
There are some examples of the processing of other types of information items during crises, including short messages (SMS)~\cite{Melville:2013:AVY:2487575.2488216}, news articles in traditional news media and blogs~\cite{leetaru_2013_gdelt}, and images~\cite{abel2012semantics}.

\spara{Verifying information.}
Systems that receive user-generated content are always exposed to abuse, which can be countered by a mixture of automatic and manual methods. For instance, algorithms to detect false product reviews are deployed by most major online retailers~\cite[Chapter 10]{liu_2012_sentiment}.
In the crisis domain, determining the credibility of information posted on social media is a major concern for those who process information (e.g. computer scientists, and software engineers), and for the information consumers (e.g. the public and formal response organizations)~\cite{hiltz2011introduction,hughes_2014_smem}.

Automatic classification can be used to filter out content that is unlikely to be considered credible~\cite{gupta2012credibility,castillo2013predicting}, or to annotate messages seen by users with credibility scores automatically~\cite{gupta2014tweetcred}. Additionally, the public itself can be mobilized to confirm or discredit a claim through crowdsourcing~\cite{popoola2013information}.

\subsection{Beyond Data Processing}

\spara{Designing with the users.}
Considering the number of systems that have been designed and built so far, there is little research on how usable and useful are those systems (with some exceptions, e.g.~\cite{robinson2013understanding,tucker2012straight}).

How should information be presented to users? How should users interact with it?
The key to answering these question lies with the users themselves, who should be brought into the process of designing the systems, dashboards, and/or visualizations they require to serve their needs. A highly regarded methodology for achieving this is \emph{participatory design}~\cite{hughes_2014_pio}.

\spara{Helping governments and NGOs communicate with the public.}
Three days \emph{before} Typhoon Pablo made its landfall in the Philippines in 2012, government officials were already calling users to use the hashtag ``\#pabloph'' for updates about the typhoon.
An effective use of hashtags has also been encouraged by the United Nations Office for the Coordination of Humanitarian Affairs~\cite{ocha_2014_hashtags}. Computational methods can be used not only to help formal response agencies choose which hashtags to use, but more generally, to help them design and evaluate effective communication strategies in social media (see~\citeN{veil2011work} regarding best practices for crisis communications using social media).
%

The conversation between the public and formal organizations can also be conducted through platforms that, instead of passively waiting for people to post information, ask them directly to answer certain questions that are relevance for the emergency response or relief operations~\cite{ludwig_2015_crowdmonitor}.

\spara{Social media for coordinating actions.}
The final output of the processing of social media messages is not limited to the presentation of information in a given format. Computational methods can be applied to augment the information in a number of ways.
For instance, Varga et al.~\citeyear{varga2013aid} match problem-tweets (``infant formula is sold out in Sendai'') to solution-tweets (``Jusco supermarket is selling infant formula''), and Purohit et al.~\citeyear{purohit2013emergency} match tweets describing urgent need of resources (``we are coordinating a clothing drive for families affected'') with tweets describing the intention to donate them (``I've a bunch of clothes I want to donate'').
We regard these efforts as preliminary results towards the ability to use social media as a mechanism for coordination of action in future emergency situations.
A recent special issue of the Journal on Computer Supported Collaborative Work explores various ways in which computing can support collaboration and coordination during an emergency~\cite{pipek_2014_special}.

\spara{Acknowledgments.} We are thankful to Jakob Rogstadius and Per Aarvik who pointed us to historical information.
We would like to thank our collaborators and co-authors Patrick Meier, Alexandra Olteanu, and Hemant Purohit.

\footnotesize{\bibliographystyle{ACM-Reference-Format-Journals}
\bibliography{references}
}
  


\end{document}